\pdfoutput=1

\documentclass[11pt]{article}

\usepackage{acl}

\usepackage{times}
\usepackage{latexsym}
\usepackage{xspace}

\usepackage[T1]{fontenc}

\usepackage[utf8]{inputenc}

\usepackage{microtype}

\usepackage{inconsolata}
\usepackage{url}            
\usepackage{booktabs}       
\usepackage{amsfonts}       
\usepackage{nicefrac}       
\usepackage{microtype}      
\usepackage{amsmath}
\usepackage{amssymb}
\usepackage{enumitem}
\usepackage{graphicx}       
\usepackage{caption}
\usepackage{subcaption}
\usepackage{pifont}
\usepackage{multirow}
\usepackage{textcomp}
\usepackage{listings}
\usepackage{flushend}
\usepackage[hang,flushmargin]{footmisc}
\usepackage{adjustbox}

\newcommand\ignore[1]{}

\title{Operational Advice for Dense and Sparse Retrievers:\\ HNSW, Flat, or Inverted Indexes?}

\author{Jimmy Lin \\[1ex]
        David R. Cheriton School of Computer Science \\
        University of Waterloo \\[1ex]
        \texttt{jimmylin@uwaterloo.ca}}

\begin{document}
\maketitle
\begin{abstract}
Practitioners working on dense retrieval today face a bewildering number of choices.
Beyond selecting the embedding model, another consequential choice is the actual implementation of nearest-neighbor vector search.
While best practices recommend HNSW indexes, flat vector indexes with brute-force search represent another viable option, particularly for smaller corpora and for rapid prototyping.
In this paper, we provide experimental results on the BEIR dataset using the open-source Lucene search library that explicate the tradeoffs between HNSW and flat indexes (including quantized variants) from the perspectives of indexing time, query evaluation performance, and retrieval quality.
With additional comparisons between dense and sparse retrievers, our results provide guidance for today's search practitioner in understanding the design space of dense and sparse retrievers.
To our knowledge, we are the first to provide operational advice supported by empirical experiments in this regard.
\end{abstract}

\section{Introduction}

Retrieval-augmented generation (RAG), which involves injecting search results into the prompt of a large language model (LLM) to provide context or ``grounding'', is one of the most popular and effective generative AI techniques today~\cite{LewisPatrick_etal_NeurIPS2020,Gao:2312.10997:2024}.
It is widely recognized that the quality of the generated responses depends to a large extent on the quality of the search results, i.e., ``garbage in, garbage out''.
This makes retrieval a critical component of RAG.

Today, practitioners typically take advantage of vector search to generate search results, but they face a bewildering number of choices.
There's first-stage retrieval to generate a list of candidates, possibly followed by reranking.
Even focused on the first stage, dense retrieval models and sparse retrieval models compete for attention, often confusing newcomers; and this is only considering single-vector variants, leaving aside multi-vector techniques such as ColBERT~\cite{Khattab_Zaharia_SIGIR2020}.
To offer a conceptual structure, \citet{Lin_arXiv2021_repir} provides a framework for thinking about retrieval in terms of nearest-neighbor search over vector representations (of queries and documents), where these representations can be dense (typically called embeddings, generated from transformers) or sparse (also generated by transformers).
Relevance is captured by simple vector operations such as the dot product, and a retriever's task is to efficiently produce the top-$k$ documents from a corpus based on these similarity comparisons.

The focus of most efforts today lie in dense retrieval models~\cite{dpr}, where queries and documents are represented by dense vectors (i.e., embeddings), typically generated by transformer models that have been fine-tuned on human-labeled or synthetically generated relevance data.
This forms the starting point of our work.
Nearest-neighbor search over dense representation vectors defines rankings of documents with respect to queries, but says nothing about how those rankings are computed efficiently at scale.
Presently, best practices recommend the use of hierarchical navigable small-world network (HNSW) indexes~\cite{HNSW}.
An alternative is so-called flat indexes that take advantage of brute-force search, which are attractive in certain scenarios.
But when?
More broadly, a search practitioner today faces choices between dense retrieval models and sparse retrieval models.
How do they navigate these options?

This work attempts to sort through these myriad options for dense and sparse retrievers, in particular focusing on three research questions:

\begin{itemize}[leftmargin=1.0cm]

\item [{\bf RQ1}] For dense retrieval, when should HNSW indexes be used vs.\ flat indexes and what are the associated tradeoffs?

\item [{\bf RQ2}] For both HNSW and flat indexes, when should quantization be applied and what are the associated tradeoffs?

\item [{\bf RQ3}] More broadly, what are the effectiveness--efficiency tradeoffs between dense and sparse retrieval across different corpora?

\end{itemize}

\noindent Ultimately, our goal is to provide operational guidance for a search practitioner to navigate the complex design space of dense and sparse retrieval.
Our goal is to explicate the tradeoffs involving indexing time, query evaluation performance, and retrieval quality to help practitioners make better decisions informed by experimental evidence.

\section{The State of the Art}
\label{section:art}

It makes sense to begin with a characterization of the state of the art, not in the sense of leaderboard chasing, but the day-to-day choices faced by search practitioners ``in the real world''.
Naturally, it is not possible to cover {\it all} aspects of retrieval, so we focus on the three main research questions articulated in the introduction.

Brute-force search with flat indexes was introduced in Elasticsearch v8.13 (released March 2024).
As Elasticsearch is built on the Lucene search library used in our experiments, it provides an appropriate starting point for our discussions.
An official blog post\footnote{\url{https://www.elastic.co/blog/whats-new-elasticsearch-platform-8-13-0}} accompanying the release offers the following advice:\ ``when the size of the set\ldots\ is rather small, it is usually better to rely on brute-force vector search rather than on HNSW-based vector search.''
But what does ``rather small'' mean?
And what other factors matter?
This advice cannot be easily operationalized, making it unhelpful for search practitioners (RQ1).
Elsewhere, we find advocates for flat indexes using DataFrames, or even Numpy,\footnote{\url{https://x.com/softwaredoug/status/1802433164201415000}} especially for rapid prototyping.
The same Elasticsearch blog post discusses int8 quantization, but is similarly vague about specific guidance (RQ2).
Finally, ``heads up'' fair comparisons between dense retrievers and alternative models are difficult to find (RQ3).

While we point to this specific instance to illustrate a gap in the state of the art, the general sentiments expressed in the blog post are not unique.
Other documents found on the web and on social media are similarly handy-wavy in providing guidance, and what few specifics offered are unsupported by empirical evidence.
To our knowledge, the concrete advice offered in this paper using an existing, widely adopted benchmark does not exist anywhere else, and represents the contribution of our work.
Of course, specific application deployments require balancing many competing factors, and it is impossible to offer ``one-size-fits-all'' advice.
Nevertheless, we provide empirical evidence that accurately characterizes the design space to inform system builders.

It is obvious that performance is affected by scale (e.g., size of corpora and length of individual documents), the embedding model, the types of queries, as well as many other factors, but it would be desirable to provide search practitioners today more specific guidance.
According to a talk by Chroma, a vector database vendor,\footnote{\url{https://www.youtube.com/watch?v=E4ot5d79jdA}} most of their customers manage corpora ranging from ``several hundreds of thousands'' to ``several millions'' vectors.
This is consistent with other discussions on social media, and provides us a point of calibration.
We structure our study in terms of corpora in this range of sizes to benefit the broadest audience.

\section{Methods}
\label{section:methods}

We begin by describing and justifying our experimental setup.
All experiments in this paper take advantage of BEIR~\cite{thakur:2021}, which comprises a large collection of individual retrieval datasets and has emerged as the standard benchmark for evaluating retrieval applications.
We provide detailed experimental results over 29 different individual datasets,\footnote{Note that CQADupStack is actually comprised of 12 different ``verticals''.} each with different corpora, queries, and task definitions.
This variety provides a cross section of search tasks and realistically reflects real-world scenarios.

Our evaluations were conducted with the open-source Lucene search library, a choice that deserves some discussion and justification.
We provide two main reasons:\
First, Lucene is the most widely deployed search library in the world, mostly via platforms such as Elasticsearch, Solr, and OpenSearch.
\citet{Devins_etal_WSDM2022} have shown that implementations in Lucene simplify many aspects of IR experiments, but yet can be easily ported over to Elasticsearch---this combination facilitates prototyping while preserving fidelity to real-world scenarios.
Thus, our results would be of broad interest to many practitioners in the community.

Second, our work with Lucene provides a comparison across dense and sparse techniques that is as fair as possible given currently available software.
While Lucene provides a production-grade implementation of HNSW indexes, it is one of many existing options currently available on the market.
Faiss~\cite{faiss} is another popular option, and there is a vibrant ecosystem of vendors providing vector search capabilities (Weaviate, Chroma, Pinecone, Vectara, and many others).
Vector search has also been integrated into relational databases~\cite{Xian_etal_WSDM2024}, for example, pgvector for Postgres.

However, we selected Lucene because it provides implementations of {\it both} dense and sparse retrieval, making comparisons reasonably fair.
For example, comparing Faiss HNSW indexes (implemented in C++) with Lucene inverted indexes (implemented in Java) or even Numpy would be conflating too many non-relevant factors (e.g., language choice).
Within the same project (Lucene), we expect different retrieval techniques to have comparable implementation quality.
While Vespa does provide dense and sparse vector search capabilities, it remains niche and lacks the wide install base of Lucene, making results of limited interest to the broader community.

\paragraph{Retrieval models.}
We examined the following retrieval models in this study:

\begin{itemize}[leftmargin=*]

\item BGE~\cite{Xiao:2309.07597:2024} was selected as a representative dense retrieval model; specifically \texttt{\small bge-base-en-v1.5}.

\item SPLADE++ EnsembleDistil (ED)~\cite{splade} was selected as a representative sparse retrieval model.

\item BM25~\cite{robertson2009bm25} provides the baseline.
For BEIR,~\citet{Kamalloo_etal_SIGIR2024} identified two distinct BM25 variants:\ here we use the approach where all document fields are concatenated prior to indexing.

\end{itemize}

\noindent For the dense retrieval model (BGE), our work examined two types of indexes.
First, we considered hierarchical navigable small-world network (HNSW) indexes~\cite{HNSW}, which represent best practices today for nearest-neighbor search over dense vectors.
Most ``vector DB'' vendors today offer variants of such indexes.

Alternatively, we evaluated so-called ``flat'' indexes, where the dense vectors are simply stored sequentially, one after the other.
``Indexing'' in this case is simply rewriting the embedding vectors in an internal representation.
Top-$k$ retrieval is implemented as brute-force search:\ the retriever simply scans the vectors, computing (in our case) the dot product between the query and each document vector, retaining only the top $k$ results.

For SPLADE++ ED, we used standard inverted indexes, taking advantage of the widely known ``fake words'' trick, where quantized impact scores replace the term frequency component in the postings, and query evaluation uses a ``sum of term frequencies'' scoring function.
See~\citet{Mackenzie_etal_TOIS2023} for more details.
BM25 also used standard inverted indexes.

\paragraph{Implementation details.}
All experiments were conducted using the Anserini open-source IR toolkit~\cite{Yang_etal_JDIQ2018}, based on Lucene 9.9.1 (released Dec.\ 2023).
We used bindings for Lucene HNSW indexes recently introduced in~\citet{Ma_etal_CIKM2023}.
We set the HNSW indexing parameters \texttt{\small M} to 16 and \texttt{\small efC} to 100, both representing typical configurations.
Lucene's HNSW indexing implementation generates different index segments and then merges them as needed in a hierarchical manner; we used all default settings here.
On the retrieval end, we set \texttt{\small efSearch} to 1000, another common setting.
The flat index implementation in Anserini is adapted from Elasticsearch.

All experiments were performed on a circa-2022 Mac Studio with an M1 Ultra processor containing 20 cores (16 performance, 4 efficiency) and 128 GB memory, running macOS Sonoma 14.5 and OpenJDK 21.0.2.
We enabled the \texttt{\small jdk.incubator.vector} module for more efficient vector operations.
Both indexing and retrieval experiments used 16 threads.
In all cases (HNSW, flat, and inverted indexes), we retrieved 1000 hits and evaluated retrieval quality in terms of nDCG@10, per BEIR guidelines.
Query evaluation performance was measured in terms of queries per second (QPS) using 16 threads.

\section{Experimental Results}

\begin{table*}[t]
\centering
\scalebox{0.78}{
\begin{tabular}{lrr|rrr|rrr|rrr|r}
\toprule
 &  &  & \multicolumn{3}{l|}{nDCG@10} & \multicolumn{3}{l|}{QPS (cached)} & \multicolumn{3}{l|}{QPS (ONNX)} & QPS \\
Dataset & $|\mathcal{C}|$ & $|Q|$ & Dense & Sparse & BM25 & {\small Flat} & {\small HNSW} & {\small INV} & {\small Flat} & {\small HNSW} & {\small INV} & BM25 \\
\midrule
NFCorpus & 3,633 & 323 & 0.373 & 0.347 & 0.322 & 270 & 280 & 430 & 210 & 200 & 220 & 480 \\
SciFact & 5,183 & 300 & 0.741 & 0.704 & 0.679 & 260 & 260 & 280 & 200 & 190 & 140 & 280 \\
ArguAna & 8,674 & 1,406 & 0.636 & 0.520 & 0.397 & 440 & 430 & 320 & 240 & 260 & 23 & 360 \\
CQA Mathematica & 16,705 & 804 & 0.316 & 0.238 & 0.202 & 330 & 340 & 350 & 240 & 240 & 210 & 390 \\
CQA webmasters & 17,405 & 506 & 0.406 & 0.317 & 0.306 & 320 & 330 & 290 & 210 & 220 & 180 & 340 \\
CQA Android & 22,998 & 699 & 0.507 & 0.390 & 0.380 & 310 & 320 & 350 & 220 & 220 & 190 & 380 \\
SCIDOCS & 25,657 & 1,000 & 0.217 & 0.159 & 0.149 & 290 & 330 & 330 & 240 & 230 & 190 & 190 \\
CQA programmers & 32,176 & 876 & 0.424 & 0.340 & 0.280 & 340 & 390 & 350 & 220 & 230 & 200 & 390 \\
CQA GIS & 37,637 & 885 & 0.413 & 0.315 & 0.290 & 350 & 360 & 380 & 220 & 230 & 190 & 380 \\
CQA physics & 38,316 & 1,039 & 0.472 & 0.360 & 0.321 & 360 & 360 & 410 & 220 & 230 & 200 & 420 \\
CQA English & 40,221 & 1,570 & 0.486 & 0.408 & 0.345 & 400 & 430 & 440 & 230 & 240 & 200 & 480 \\
CQA stats & 42,269 & 652 & 0.373 & 0.299 & 0.271 & 290 & 310 & 350 & 200 & 210 & 180 & 340 \\
CQA gaming & 45,301 & 1,595 & 0.597 & 0.496 & 0.482 & 410 & 430 & 430 & 230 & 240 & 210 & 460 \\
CQA UNIX & 47,382 & 1,072 & 0.422 & 0.317 & 0.275 & 360 & 360 & 410 & 210 & 230 & 200 & 390 \\
CQA Wordpress & 48,605 & 541 & 0.355 & 0.273 & 0.248 & 310 & 350 & 310 & 190 & 200 & 180 & 320 \\
FiQA-2018 & 57,638 & 648 & 0.406 & 0.347 & 0.236 & 290 & 330 & 300 & 190 & 220 & 170 & 340 \\
CQA tex & 68,184 & 2,906 & 0.311 & 0.253 & 0.224 & 400 & 480 & 520 & 210 & 240 & 220 & 490 \\
\midrule
TREC-COVID & 171,332 & 50 & 0.781 & 0.727 & 0.595 & 66 & 100 & 65 & 58 & 76 & 52 & 92 \\
Touch\'{e} 2020 & 382,545 & 49 & 0.257 & 0.247 & 0.442 & 38 & 85 & 52 & 33 & 61 & 47 & 68 \\
Quora & 522,931 & 10,000 & 0.889 & 0.834 & 0.789 & 75 & 200 & 420 & 61 & 110 & 180 & 770 \\
Robust04 & 528,155 & 249 & 0.447 & 0.468 & 0.407 & 57 & 150 & 150 & 48 & 89 & 86 & 110 \\
TREC-NEWS & 594,977 & 57 & 0.443 & 0.415 & 0.395 & 29 & 72 & 54 & 27 & 67 & 47 & 47 \\
\midrule
NQ & 2,681,468 & 3,452 & 0.541 & 0.538 & 0.305 & 15 & 140 & 130 & 14 & 90 & 85 & 470 \\
Signal-1M & 2,866,316 & 97 & 0.289 & 0.301 & 0.330 & 8.8 & 60 & 59 & 8.5 & 41 & 46 & 180 \\
DBpedia & 4,635,922 & 400 & 0.407 & 0.437 & 0.318 & 7.7 & 72 & 80 & 7.4 & 52 & 63 & 300 \\
HotpotQA & 5,233,329 & 7,405 & 0.726 & 0.687 & 0.633 & 7.6 & 74 & 69 & 7.4 & 52 & 46 & 460 \\
FEVER & 5,416,568 & 6,666 & 0.863 & 0.788 & 0.651 & 7.3 & 63 & 65 & 7.2 & 47 & 49 & 470 \\
Climate-FEVER & 5,416,593 & 1,535 & 0.312 & 0.230 & 0.165 & 7.1 & 62 & 73 & 6.9 & 44 & 47 & 290 \\
BioASQ & 14,914,603 & 500 & 0.415 & 0.498 & 0.522 & 2.6 & 56 & 24 & 2.6 & 40 & 23 & 210 \\

\bottomrule
\end{tabular}%
}
\caption{Main results comparing flat and HNSW indexes (BGE) and inverted indexes (SPLADE and BM25) in terms of effectiveness (nDCG@10) and query evaluation performance (queries per second, QPS). For nDCG@10, ``Dense'' refers to BGE and ``Sparse'' refers to SPLADE; ``INV'' refers to inverted indexes.}
\label{table:main-results}
\end{table*}

We begin with a comparison between flat, HNSW, and inverted indexes in terms of effectiveness and efficiency, shown in Table~\ref{table:main-results}.
Each row captures a dataset from BEIR, with CQADupStack broken down into its components.
The rows are sorted by the size of each corpus (number of documents, $|\mathcal{C}|$), so scanning down the rows, we encounter datasets of increasing size.
The table is informally divided into three sections that we characterize as ``small'' (less then 100K documents), ``medium'' (between 100K and 1M), and ``large'' (more than 1M).
The column marked $|Q|$ shows the number of queries in each dataset; this is important because performance measurements on smaller query sets are noisier.
The next three columns show the effectiveness of the dense model (BGE), the sparse model (SPLADE), and BM25.

Query evaluation performance is captured in terms of queries per second (QPS).
Due to the inherent noise in these measurements, we only report figures to two significant digits because any addition precision is unlikely to be meaningful.
Our experiments are divided into two conditions, cached queries and ``on-the-fly'' query encoding using ONNX (not applicable to BM25).
With cached queries, we are \textit{not} measuring the latency associated with query encoding, whereas with ONNX, latency includes query encoding.
These two measurements bookend the performance range:\ our ONNX encoding is performed on the CPU, and hence can be accelerated with GPU inference, but performance cannot exceed the cached condition.
More details about ONNX integration in Anserini are discussed in~\citet{Chen_etal_arXiv2023}.

In terms of realism:\ Obviously, in production settings, query evaluation performance must necessarily include query encoding, as the system does not know the queries in advance.
However, in a prototyping setting, or when running benchmarks repeatedly, it makes sense to cache the query representations.
Thus, we believe that both ways of measuring performance are informative, but for different scenarios.

\begin{table*}[t]
\centering
\scalebox{0.78}{
\begin{tabular}{lr|rr|rrlr}
\toprule
 &  & \multicolumn{2}{c|}{Index Time} & \multicolumn{4}{l}{nDCG@10} \\
Dataset & $|\mathcal{C}|$ & Flat & HNSW &  & avg $\Delta$ & \multicolumn{1}{c}{min} & \multicolumn{1}{c}{max} \\
\midrule
TREC-COVID & 171,332 & 0.9 & 1.8 & 0.781 & 0.000 & $[$ 0.000 & 0.000 $]$ \\
Touch\'{e} 2020 & 382,545 & 1.0 & 1.9 & 0.257 & 0.000 & $[$ 0.000 & 0.000 $]$ \\
Quora & 522,931 & 1.0 & 2.4 & 0.889 & 0.000 & $[$ 0.000 & 0.000 $]$ \\
Robust04 & 528,155 & 1.0 & 2.1 & 0.447 & 0.001 & $[$ 0.000 & 0.001 $]$ \\
TREC-NEWS & 594,977 & 1.0 & 2.1 & 0.443 & 0.001 & $[$ $-$0.004 & 0.007 $]$ \\
\midrule
NQ & 2,681,468 & 2.4 & 15.6 & 0.541 & 0.002 & $[$ 0.001 & 0.003 $]$ \\
Signal-1M & 2,866,316 & 2.3 & 14.5 & 0.289 & 0.010 & $[$ 0.006 & 0.013 $]$ \\
DBpedia & 4,635,922 & 3.2 & 31.5 & 0.407 & 0.001 & $[$ $-$0.001 & 0.004 $]$ \\
HotpotQA & 5,233,329 & 4.1 & 33.3 & 0.726 & 0.010 & $[$ 0.009 & 0.011 $]$ \\
FEVER & 5,416,568 & 4.2 & 35.0 & 0.863 & 0.005 & $[$ 0.004 & 0.006 $]$ \\
Climate-FEVER & 5,416,593 & 4.1 & 35.2 & 0.312 & 0.000 & $[$ 0.000 & 0.000 $]$ \\
BioASQ & 14,914,603 & 10.1 & 76.3 & 0.415 & 0.015 & $[$ 0.011 & 0.020 $]$ \\

\bottomrule
\end{tabular}%
}
\caption{Comparing flat indexes with HNSW indexes using the BGE embedding model. Indexing times are reported in minutes. The ``avg $\Delta$'' column reports the average degradation of HNSW scores over five trials (i.e., HNSW scores are lower); ``min'' and ``max'' report the observed min and max values across the trials; negative values indicate that HNSW indexes achieved higher scores than flat indexes.}
\label{table:rq1}
\end{table*}

\subsection{Flat vs.\ HNSW Indexes}

\begin{itemize}[leftmargin=1.0cm]

\item [{\bf RQ1}] For dense retrieval, when should HNSW indexes be used vs.\ flat indexes and what are the associated tradeoffs?

\end{itemize}

\noindent Table~\ref{table:main-results} provides guidance for this research question, illustrated with the BGE model.
Most pertinent is the comparison between flat and HNSW indexes under the ``cached'' and ``ONNX'' conditions.
We make the following observations:

\begin{itemize}[leftmargin=*]

\item For ``small'' corpora less than 100K documents, there appear to be negligible differences between flat and HNSW indexes.
For example, in an exploratory or prototyping setting, we do not see the differences in QPS as meaningful.

\item For ``medium'' corpora (between 100K and 1M), the performance differences between flat indexes and HNSW indexes become larger:\ very roughly, flat indexes are 2--3$\times$ slower with cached query representations, but after factoring in query encoding (ONNX), the gap is reduced.
For a practitioner prototyping with a small set of queries, we would recommend flat indexes, since operationally, the QPS differences are likely not meaningful.
As an example, on TREC-NEWS, the wallclock difference in evaluating on the set of 57 queries is around a second at the most.

\item For ``large'' corpora (more than 1M), the performance differences can be quite large:\ flat indexes are up to an order of magnitude slower than HNSW indexes for corpora in the 2M--5M documents range, and even slower for BioASQ, the largest BEIR corpora, at $\sim$15M documents.

\end{itemize}

\noindent To more fully characterize these tradeoffs, we need to examine two additional aspects of the design space:\ indexing time and retrieval quality.
Once again, we focus on the BGE dense retrieval model.
In Table~\ref{table:rq1}, the columns ``Flat'' and ``HNSW'' compare indexing time, averaged over five trials, rounded to the nearest tenth of a minute.
Rows are sorted by increasing size, same as in Table~\ref{table:main-results}.
For brevity, we omit results for small corpora, where the indexing times are for the most part well under a minute and the results are uninteresting.

For medium corpora (under 1M documents), we argue that the differences in indexing times are not meaningful, but the differences appear to grow as the corpus size increases; for corpora with more than 1M documents, the HNSW indexing time can be several times longer.
With flat indexes, ``indexing'' simply involves reading input vectors and rewriting them in Lucene's internal representation.
On the other hand, Lucene's HNSW indexing implementation requires building traversal graphs over segments of documents and then hierarchically merging them; indexing time does not appear to be linear with respect to corpus size.

\begin{table*}[t]
\centering
\scalebox{0.78}{
\begin{tabular}{lr|rr|rr|rr|rrlr}
\toprule
&  & \multicolumn{2}{l|}{Index Time} & \multicolumn{2}{l|}{QPS (Cached)} & \multicolumn{2}{l|}{QPS (ONNX)} & \multicolumn{4}{l}{nDCG@10} \\
Dataset & $|\mathcal{C}|$ &  & $\Delta$ &  & $\Delta$ &  & $\Delta$ &  & avg $\Delta$ & \multicolumn{1}{c}{min} & \multicolumn{1}{c}{max} \\
\midrule
TREC-COVID      &    171,332 &  0.9&$\sim$ &66&$+$3.8\% &58& $\sim$     &0.781&	$-$0.003	&$[$ $-$0.003	&-0.002 $]$\\
Touch\'{e} 2020 &    382,545 &  1.0&$\sim$ &38& $+$31\% &33& $+$25\%&0.257&	0.007	&$[$ 0.006	&0.008 $]$\\
Quora           &    522,931 &  1.0&$+$6\% &75& $+$26\% &61& $+$15\%&0.889&	0.001	&$[$ 0.001	&0.001 $]$\\
Robust04        &    528,155 &  1.0&$\sim$ &57& $+$28\% &48& $+$21\%&0.447&	0.001	&$[$ $-$0.001	&0.001 $]$\\
TREC-NEWS       &    594,977 &  1.0&$\sim$ &29& $+$48\% &27& $+$48\%&0.443&	0.009	&$[$ 0.007	&0.012 $]$\\
\midrule
NQ              &  2,681,468 &  2.4&$\sim$ &15& $+$35\% &14& $+$29\%&0.541&	0.002	&$[$ 0.002	&0.003 $]$\\
Signal-1M       &  2,866,316 &  2.3& $+$10\% &8.8& $+$63\% &8.5& $+$62\%&0.289&	0.004	&$[$ 0.002	&0.006 $]$\\
DBpedia         &  4,635,922 &  3.2& $+$17\% &7.7& $+$47\% &7.4& $+$45\%&0.407&	$-$0.001	&$[$ $-$0.002	&0.000 $]$\\
HotpotQA        &  5,233,329 &  4.1& $+$14\% &7.6& $+$36\% &7.4& $+$33\%&0.726&	0.000	&$[$ 0.000	&0.000 $]$\\
FEVER           &  5,416,568 &  4.2& $+$13\% &7.3& $+$36\% &7.2& $+$33\%&0.863&	0.001	&$[$ 0.000	&0.001 $]$\\
Climate-FEVER   &  5,416,593 &  4.1& $+$15\% &7.1& $+$39\% &6.9& $+$38\%&0.312&	0.003	&$[$ 0.002	&0.004 $]$\\
BioASQ          & 14,914,603 & 10.1& $+$12\% &2.6& $+$38\% &2.6& $+$37\%&0.415&	0.003	&$[$ 0.003	&0.003 $]$\\

\bottomrule
\end{tabular}%
}
\caption{The effects of (int8) quantization for flat indexes, compared to non-quantized versions with respect to indexing time, query evaluation performance, and retrieval quality.}
\label{table:rq2-flat}
\end{table*}

\begin{table*}[t]
\centering
\scalebox{0.78}{
\begin{tabular}{lr|rr|rr|rr|rrlr}
\toprule
&  & \multicolumn{2}{l|}{Index Time} & \multicolumn{2}{l|}{QPS (Cached)} & \multicolumn{2}{l|}{QPS (ONNX)} & \multicolumn{4}{l}{nDCG@10} \\
Dataset & $|\mathcal{C}|$ &  & $\Delta$ &  & $\Delta$ &  & $\Delta$ &  & avg $\Delta$ & \multicolumn{1}{c}{min} & \multicolumn{1}{c}{max} \\
\midrule
TREC-COVID      &    171,332 & 1.8  & $\sim$ &100 & $\sim$   &  76 &    $\sim$& 0.781&$-$0.003&	$[$ $-$0.003&	$-$0.002 $]$\\
Touch\'{e} 2020 &    382,545 & 1.9  & $\sim$ & 85 & $+$6\% &  61 &  $+$11\% & 0.257&0.006&	$[$ 0.006&	0.007 $]$\\
Quora           &    522,931 & 2.4  & $\sim$ &200 &  $+$44\% & 110 &  $+$29\% & 0.889&0.001&	$[$ 0.001&	0.001 $]$\\
Robust04        &    528,155 & 2.1  & $\sim$ &150 &  $+$21\% &  89 &  $+$22\% & 0.447&0.001&	$[$ $-$0.001&	0.003 $]$\\
TREC-NEWS       &    594,977 & 2.1  & $\sim$ & 72 &  $+$22\% &  67 &    $\sim$& 0.443&0.011&	$[$ 0.009&	0.013 $]$\\
\midrule
NQ              &  2,681,468 & 15.6 & $+$33\% &140&  $+$47\% &  90 &  $+$29\% & 0.541&0.003&	$[$ 0.002&	0.004 $]$\\
Signal-1M       &  2,866,316 & 14.5 & $+$46\% & 60&  $+$57\% &  41 &  $+$63\% & 0.289&0.010&	$[$ 0.007&	0.015 $]$\\
DBpedia         &  4,635,922 & 31.5 & $+$55\% & 72&  $+$76\% &  52 &  $+$58\% & 0.407&$-$0.001&	$[$ $-$0.004&	0.000 $]$\\
HotpotQA        &  5,233,329 & 33.3 & $+$60\% & 74& $+$130\% &  52 &  $+$90\% & 0.726&0.018&	$[$ 0.016&	0.019 $]$\\
FEVER           &  5,416,568 & 35.0 & $+$73\% & 63& $+$143\% &  47 & $+$104\% & 0.863&0.010&	$[$ 0.008&	0.012 $]$\\
Climate-FEVER   &  5,416,593 & 35.2 & $+$79\% & 62& $+$142\% &  44 &  $+$98\% & 0.312&0.001&	$[$ 0.000&	0.002 $]$\\
BioASQ          & 14,914,603 & 76.3 & $+$5\%  & 56&  $+$29\% &  40 &  $+$25\% & 0.415&0.017&	$[$ 0.011&	0.024 $]$\\

\bottomrule
\end{tabular}%
}
\caption{The effects of (int8) quantization for HNSW indexes, compared to non-quantized versions with respect to indexing time, query evaluation performance, and retrieval quality. Note that exact rankings from flat indexes provide the reference nDCG@10 scores.}
\label{table:rq2-hnsw}
\end{table*}

The retrieval quality (effectiveness) implications of flat vs.\ HNSW indexes using the BGE embedding model are also shown in Table~\ref{table:rq1}, in the columns grouped under nDCG@10.
The scores are the same as in Table~\ref{table:main-results}, under the ``Dense'' column.
Flat indexes, which yield exact similarity scores, provide the ground truth reference.
Since HNSW indexes enable fast {\it approximate} nearest-neighbor search, there is typically some effectiveness degradation, i.e., scores from HNSW indexes are usually lower.
Furthermore, since HNSW index construction is non-deterministic, scores from each trial may differ slightly.
The ``avg $\Delta$'' column reports the average degradation of HNSW scores over five trials.
The ``min'' and ``max'' columns report the observed min and max values across the trials; negative values indicate that a particular HNSW trial achieved a higher score than the corresponding flat index (sometimes possible).
 
Tables~\ref{table:main-results} and~\ref{table:rq1} together characterize the tradeoffs between flat and HNSW indexes.
For ``medium'' corpora, HNSW indexing is slower than flat indexing, but we argue that the differences are not meaningful.
There are also some effectiveness differences, but they are mostly small.
For ``large'' corpora (more than 1M documents), we see interesting tradeoffs in indexing time versus query evaluation performance.
The much higher QPS we report in Table~\ref{table:main-results} comes at a large cost in indexing time; HNSW indexes can take much longer to build than flat indexes.
Also, we observe that retrieval quality degrades more as corpus size increases.

\subsection{The Impact of Quantization}

\begin{itemize}[leftmargin=1.0cm]

\item [{\bf RQ2}] For both HNSW and flat indexes, when should quantization be applied and what are the associated tradeoffs?

\end{itemize}

\noindent Here, we examine flat and HNSW indexes separately.
Results comparing flat and quantized (int8) flat indexes are reported in Table~\ref{table:rq2-flat}.
Our analysis is organized into three relevant factors, as before:\ indexing time, query evaluation performance (QPS), and retrieval quality (nDCG@10).
Note that index quantization in Lucene is {\it not} deterministic, and we report figures averaged across five trials.
The reference indexing times for flat indexes are copied from Table~\ref{table:rq1} (measured in minutes), with $\Delta$ reporting the increase in indexing time due to quantization (as a percentage).
Similarly, query performance (QPS) under the cached and ONNX conditions are copied from Table~\ref{table:main-results} for the reference (non-quantized) condition:\ the $\Delta$ columns show increases in QPS from quantization.
In the table, $\sim$ refers to differences less than 5\%, since our measurements are noisy and we do not wish to draw attention to small differences that are likely not meaningful.
Overall, we see that quantization provides a big boost in performance (QPS) at a relatively low cost in additional indexing time.

Finally, nDCG@10 differences are organized in the same way as in Table~\ref{table:rq1}, where we report average, min, and max with respect to (non-quantized) flat indexes.
Negative values indicate that quantization {\it increased} effectiveness (possible in some cases).
Nevertheless, quantization has a relatively minor impact on retrieval quality overall.

Results comparing HNSW and quantized (int8) HNSW indexes are reported in Table~\ref{table:rq2-hnsw}, which is organized in the same manner as Table~\ref{table:rq2-flat}.
Note, however, that the reference nDCG@10 scores here are taken from exact rankings using flat indexes.
This means that the measure of degradation includes {\it both} HNSW indexing and quantization.

For HNSW indexes, we observe quantization tradeoffs that are different from flat indexes.
With medium corpora, there does not appear to be meaningful increases in indexing time, but with large corpora, indexing appears to be much slower.
Interestingly, for BioASQ, the increase in indexing time is only marginal,\footnote{Nope, verified that this isn't a bug.} which suggests that the additional costs associated with quantization are masked by other components of the indexing pipeline.

Quantization for HNSW indexes, however, delivers large benefits in increased QPS, even more than for quantization in flat indexes.
The effectiveness degradation of quantized HNSW indexes is comparable to non-quantized HNSW indexes, which suggests that the effectiveness impact of quantization is minor at most.

\subsection{Dense Retrieval in a Broader Context}

\begin{itemize}[leftmargin=1.0cm]

\item [{\bf RQ3}] More broadly, what are the effectiveness--efficiency tradeoffs between dense and sparse retrieval across different corpora?

\end{itemize}

\noindent Effectiveness comparisons of dense and sparse retrieval models abound in the literature and are found in almost every retrieval modelling paper published today.
Overall, one approach does not appear to be dominant, and it might be fair to characterize dense and sparse models as comparable in terms of effectiveness.

However, query evaluation performance has received little attention by researchers, and we contribute a comparison between SPLADE++ ED and BGE in a fair setting, shown in Table~\ref{table:main-results}.
In terms of QPS, both appear to be comparable, looking at the ``HNSW'' vs.\ ``INV'' columns.\footnote{ArguAna appears to be an outlier for SPLADE; we verified that this was not a bug.}
There does not appear to be a compelling reason to choose dense retrieval over sparse retrieval (or vice versa) from the performance point of view.
Indeed, the literature is consistent in advocating hybrid techniques that combine both approaches~\cite{thakur:2021,Ma_etal_ECIR2022,Kamalloo_etal_SIGIR2024}.

Table~\ref{table:main-results} also provides a comparison between SPLADE++ ED and BM25.
In terms of effectiveness, the SPLADE model dominates BM25 and outperforms it for nearly all of the datasets in BEIR; the exceptions are Touch\'{e}, Signal-1M, and BioASQ.
In the first case,~\citet{Thakur_etal_SIGIR2024} provides a detailed error analysis explaining these results.
However, we see from the final column that BM25 is much faster than SPLADE++ ED; the difference is close to an order of magnitude in the case of BioASQ, the largest corpus.
For some points in the effectiveness--efficiency tradeoff space, there is still a role for BM25.

\section{Discussion}

The primary goal of this paper is to replace ``hand waving'' with empirical evidence for the benefit of search practitioners.
Our experimental results illustrate the tradeoff space with BEIR, a widely adopted retrieval benchmark.
While the ultimate choices of system builders will depend on the real-world scenario (from prototyping to proof of concept to production deployment), we can offer some advice.
At a high level, for corpora with fewer than 1M documents, we do not see a compelling advantage to using HNSW indexes.
For larger corpora, however, we feel that the advantages of HNSW indexes in terms of query evaluation performance offset the downsides.

Another issue worth explicitly discussing is the retrieval quality degradation associated with HNSW indexing and quantization.
These factors are not typically discussed in academic research, but are important from the perspective of building real-world systems.
A recap of the issues:\ both HNSW indexing and quantization are non-deterministic and typically degrade retrieval effectiveness with respect to exact similarity comparisons (captured in flat indexes).
As an example, BioASQ results from Table~\ref{table:rq2-hnsw} show that, with HNSW and quantization, nDCG@10 scores are 0.017 lower (averaged across five trials), with a max difference of 0.024; this translates into $4\%$ and $5.8\%$, respectively---relatively large differences.
These effects are potentially problematic when comparing different embedding models that are ``close'' in terms of quality, because it would be hard to tease apart model quality from an ``unlucky'' sub-optimal index.
Nearly all academic papers sweep these differences under the rug, but they represent important practical considerations.
For this reason, flat indexes are appealing for rapid prototyping in order to isolate the quality of embedding models.

\section{Conclusions}

There are three main limitations to this work worth pointing out.
First, we study only a single instance of a dense and sparse retrieval model (BGE and SPLADE++ ED).
While both are popular and representative, there are many other models worth considering and new ones appearing all the time.
Second, we only evaluate performance on a single system.
An exhaustive matrix experiment involving different models and systems (architectures, OSes, etc.)\ would be impractical, and we expect the broad contours of our findings to remain invariant.
However, more experiments are needed to confirm the generalizability of our findings.

Another limitation is our focus on Lucene, even though there are many other HNSW implementations available.
This issue has already been discussed in Section~\ref{section:methods}, and it may be the case that other system combinations will alter our conclusions.
However, as we pointed out, such comparisons are difficult to set up in a fair manner.
Nevertheless, the dominance of Lucene means that our findings are of broad interest, worthy of consideration even for users of other platforms.

There are many more decisions that a search practitioner needs to make when building a full RAG system, beyond the explicit research questions that we consider in this work.
For example, what are the roles of reranking and prompt engineering?
How do we deal with dynamically changing documents?
The list goes on.
Nevertheless, we hope that this work offers a starting point in providing empirically grounded guidance for search practitioners building real-world applications.

\section*{Acknowledgements}

This research was supported in part by the Natural Sciences and Engineering Research Council (NSERC).
We'd like to acknowledge Snowflake for additional funding.
Thanks to Steven Chen for helpful comments on an earlier draft of this paper.

\bibliography{custom}

\begin{thebibliography}{20}
\expandafter\ifx\csname natexlab\endcsname\relax\def\natexlab#1{#1}\fi

\bibitem[{Chen et~al.(2023)Chen, Lassance, and Lin}]{Chen_etal_arXiv2023}
Haonan Chen, Carlos Lassance, and Jimmy Lin. 2023.
\newblock End-to-end retrieval with learned dense and sparse representations
  using {Lucene}.
\newblock \emph{arXiv:2311.18503}.

\bibitem[{Devins et~al.(2022)Devins, Tibshirani, and
  Lin}]{Devins_etal_WSDM2022}
Josh Devins, Julie Tibshirani, and Jimmy Lin. 2022.
\newblock Aligning the research and practice of building search applications:
  {Elasticsearch} and {Pyserini}.
\newblock In \emph{Proceedings of the 15th ACM International Conference on Web
  Search and Data Mining (WSDM 2022)}, pages 1573--1576.

\bibitem[{Formal et~al.(2022)Formal, Lassance, Piwowarski, and
  Clinchant}]{splade}
Thibault Formal, Carlos Lassance, Benjamin Piwowarski, and St{\'e}phane
  Clinchant. 2022.
\newblock From distillation to hard negative sampling: Making sparse neural
  {IR} models more effective.
\newblock In \emph{Proceedings of the 45th International ACM SIGIR Conference
  on Research and Development in Information Retrieval}, pages 2353--2359.

\bibitem[{Gao et~al.(2024)Gao, Xiong, Gao, Jia, Pan, Bi, Dai, Sun, Wang, and
  Wang}]{Gao:2312.10997:2024}
Yunfan Gao, Yun Xiong, Xinyu Gao, Kangxiang Jia, Jinliu Pan, Yuxi Bi, Yi~Dai,
  Jiawei Sun, Meng Wang, and Haofen Wang. 2024.
\newblock Retrieval-augmented generation for large language models: A survey.
\newblock \emph{arXiv:2312.10997}.

\bibitem[{Johnson et~al.(2019)Johnson, Douze, and J{\'e}gou}]{faiss}
Jeff Johnson, Matthijs Douze, and Herv{\'e} J{\'e}gou. 2019.
\newblock Billion-scale similarity search with {GPUs}.
\newblock \emph{IEEE Transactions on Big Data}, 7(3):535--547.

\bibitem[{Kamalloo et~al.(2024)Kamalloo, Thakur, Lassance, Ma, Yang, and
  Lin}]{Kamalloo_etal_SIGIR2024}
Ehsan Kamalloo, Nandan Thakur, Carlos Lassance, Xueguang Ma, Jheng-Hong Yang,
  and Jimmy Lin. 2024.
\newblock Resources for brewing {BEIR}: Reproducible reference models and
  statistical analyses.
\newblock In \emph{Proceedings of the 47th International ACM SIGIR Conference
  on Research and Development in Information Retrieval (SIGIR 2024)}, pages
  1431--1440, Washington, D.C.

\bibitem[{Karpukhin et~al.(2020)Karpukhin, Oguz, Min, Lewis, Wu, Edunov, Chen,
  and Yih}]{dpr}
Vladimir Karpukhin, Barlas Oguz, Sewon Min, Patrick Lewis, Ledell Wu, Sergey
  Edunov, Danqi Chen, and Wen-tau Yih. 2020.
\newblock Dense passage retrieval for open-domain question answering.
\newblock In \emph{Proceedings of the 2020 Conference on Empirical Methods in
  Natural Language Processing (EMNLP)}, pages 6769--6781, Online.

\bibitem[{Khattab and Zaharia(2020)}]{Khattab_Zaharia_SIGIR2020}
Omar Khattab and Matei Zaharia. 2020.
\newblock {ColBERT}: Efficient and effective passage search via contextualized
  late interaction over {BERT}.
\newblock In \emph{Proceedings of the 43rd International ACM SIGIR Conference
  on Research and Development in Information Retrieval (SIGIR 2020)}, pages
  39--48.

\bibitem[{Lewis et~al.(2020)Lewis, Perez, Piktus, Petroni, Karpukhin, Goyal,
  {K\"{u}ttler}, Lewis, tau Yih, {Rockt\"{a}schel}, Riedel, and
  Kiela}]{LewisPatrick_etal_NeurIPS2020}
Patrick Lewis, Ethan Perez, Aleksandra Piktus, Fabio Petroni, Vladimir
  Karpukhin, Naman Goyal, Heinrich {K\"{u}ttler}, Mike Lewis, Wen tau Yih, Tim
  {Rockt\"{a}schel}, Sebastian Riedel, and Douwe Kiela. 2020.
\newblock Retrieval-augmented generation for knowledge-intensive nlp tasks.
\newblock In \emph{Advances in Neural Information Processing Systems 33}, pages
  9459--9474.

\bibitem[{Lin(2021)}]{Lin_arXiv2021_repir}
Jimmy Lin. 2021.
\newblock A proposed conceptual framework for a representational approach to
  information retrieval.
\newblock \emph{arXiv:2110.01529}.

\bibitem[{Ma et~al.(2022)Ma, Sun, Pradeep, Li, and Lin}]{Ma_etal_ECIR2022}
Xueguang Ma, Kai Sun, Ronak Pradeep, Minghan Li, and Jimmy Lin. 2022.
\newblock Another look at {DPR}: Reproduction of training and replication of
  retrieval.
\newblock In \emph{Proceedings of the 44th European Conference on Information
  Retrieval (ECIR 2022), Part I}, pages 613--626, Stavanger, Norway.

\bibitem[{Ma et~al.(2023)Ma, Teofili, and Lin}]{Ma_etal_CIKM2023}
Xueguang Ma, Tommaso Teofili, and Jimmy Lin. 2023.
\newblock {Anserini} gets dense retrieval: Integration of {Lucene's} {HNSW}
  indexes.
\newblock In \emph{Proceedings of the 32nd International Conference on
  Information and Knowledge Management (CIKM 2023)}, pages 5366--5370,
  Birmingham, the United Kingdom.

\bibitem[{Mackenzie et~al.(2022)Mackenzie, Trotman, and
  Lin}]{Mackenzie_etal_TOIS2023}
Joel Mackenzie, Andrew Trotman, and Jimmy Lin. 2022.
\newblock Efficient document-at-a-time and score-at-a-time query evaluation for
  learned sparse representations.
\newblock \emph{ACM Transactions on Information Systems}, 41:Article No.~96.

\bibitem[{Malkov and Yashunin(2020)}]{HNSW}
Yu~A. Malkov and D.~A. Yashunin. 2020.
\newblock Efficient and robust approximate nearest neighbor search using
  hierarchical navigable small world graphs.
\newblock \emph{Transactions on Pattern Analysis and Machine Intelligence},
  42(4):824--836.

\bibitem[{Robertson and Zaragoza(2009)}]{robertson2009bm25}
Stephen~E. Robertson and Hugo Zaragoza. 2009.
\newblock The probabilistic relevance framework: {BM25} and beyond.
\newblock \emph{Foundations and Trends in Information Retrieval},
  3(4):333--389.

\bibitem[{Thakur et~al.(2024)Thakur, Bonifacio, {Fr\"{o}be}, Bondarenko,
  Kamalloo, Potthast, Hagen, and Lin}]{Thakur_etal_SIGIR2024}
Nandan Thakur, Luiz Bonifacio, Maik {Fr\"{o}be}, Alexander Bondarenko, Ehsan
  Kamalloo, Martin Potthast, Matthias Hagen, and Jimmy Lin. 2024.
\newblock Systematic evaluation of neural retrieval models on the {Touch\'{e}}
  2020 argument retrieval subset of {BEIR}.
\newblock In \emph{Proceedings of the 47th International ACM SIGIR Conference
  on Research and Development in Information Retrieval (SIGIR 2024)}, pages
  1420--1430, Washington, D.C.

\bibitem[{Thakur et~al.(2021)Thakur, Reimers, R{\"{u}}ckl{\'{e}}, Srivastava,
  and Gurevych}]{thakur:2021}
Nandan Thakur, Nils Reimers, Andreas R{\"{u}}ckl{\'{e}}, Abhishek Srivastava,
  and Iryna Gurevych. 2021.
\newblock {BEIR:} {A} heterogeneous benchmark for zero-shot evaluation of
  information retrieval models.
\newblock In \emph{Proceedings of the Neural Information Processing Systems
  Track on Datasets and Benchmarks 1, NeurIPS Datasets and Benchmarks 2021}.

\bibitem[{Xian et~al.(2024)Xian, Teofili, Pradeep, and
  Lin}]{Xian_etal_WSDM2024}
Jasper Xian, Tommaso Teofili, Ronak Pradeep, and Jimmy Lin. 2024.
\newblock Vector search with {OpenAI} embeddings: {Lucene} is all you need.
\newblock In \emph{Proceedings of the 17th ACM International Conference on Web
  Search and Data Mining (WSDM 2024)}, pages 1090--1093, {M\'{e}rida},
  {M\'{e}xico}.

\bibitem[{Xiao et~al.(2024)Xiao, Liu, Zhang, Muennighoff, Lian, and
  Nie}]{Xiao:2309.07597:2024}
Shitao Xiao, Zheng Liu, Peitian Zhang, Niklas Muennighoff, Defu Lian, and
  Jian-Yun Nie. 2024.
\newblock {C-Pack}: Packaged resources to advance general {Chinese} embedding.
\newblock \emph{arXiv:2309.07597}.

\bibitem[{Yang et~al.(2018)Yang, Fang, and Lin}]{Yang_etal_JDIQ2018}
Peilin Yang, Hui Fang, and Jimmy Lin. 2018.
\newblock {Anserini}: Reproducible ranking baselines using {Lucene}.
\newblock \emph{Journal of Data and Information Quality}, 10(4):Article 16.

\end{thebibliography}

\end{document}